\documentclass[12pt]{amsart}
\usepackage{amsmath}
\usepackage{amssymb}
\usepackage{latexsym}

\newcommand{\Zint}{{\bf Z}}    
     
\newcommand{\Rea}{{\bf {R}}}      % Real number field
\newcommand{\Cplx}{{\bf {C}}}     % Complex  number field

\newtheorem{prop}{\bf Proposition}[section]
\newtheorem{thm}[prop]{\bf Theorem}

\newtheorem{cor}[prop]{\bf Corollary}

\begin{document}

\title{Heun equation and Inozemtsev models}
\author{Kouichi TAKEMURA}
\address{Department of Mathematical Sciences, Yokohama City University, 22-2 Seto, 
Kanazawa-ku, Yokohama 236-0027, JAPAN}
\begin{abstract}
The $BC_N$ elliptic Inozemtsev model is a quantum integrable systems with $N$-particles whose potential is given by elliptic functions. Eigenstates and eigenvalues of this model are investigated.
\end{abstract}
% Leave the next line commented out!
\maketitle

\section{Introduction}

The $BC_N$ Inozemtsev model \cite{Ino} is a system of quantum mechanics with $N$-particles whose Hamiltonian is given by
\begin{eqnarray}
& H=-\sum_{j=1}^N\frac{\partial ^2}{\partial x_j^2} + \sum_{j=1}^N \sum _{i=0}^3 l_i(l_i+1) \wp(x_j +\omega_i)  \label{InoHam} & \\
& \; \; \; \; \; + 2l(l+1)\sum_{1\leq j<k\leq N} \left( \wp (x_j-x_k) +\wp (x_j +x_k) \right) , \nonumber & 
\end{eqnarray}
where $\wp (x)$ is the Weierstrass $\wp$-function with periods $(1,\tau )$, $\omega _0=0$, $\omega_1=\frac{1}{2}$, $\omega_2=-\frac{\tau+1}{2}$, $\omega_3=\frac{\tau}{2}$ are half periods, and $l$ and $l_i$ $(i=0,1,2,3)$ are coupling constants.

It is known that the $BC_N$ Inozemtsev model is quantum completely integrable.
%that is a quantum version of the Liouville's integrability.
More precisely, there exist operators of the form $H_k= \sum_{j=1}^N \left( \frac{\partial }{\partial x_j} \right) ^{2k} + \mbox{(lower terms)}$ $(k=2,\dots ,N)$ such that $[H, H_k]=0$ and $[H_{k_1}, H_{k_2}]=0$ $(k, k_1, k_2=2,\dots ,N)$.
%For the $BC_N$ Inozemtsev model, Oshima \cite{O} wrote down the commuting operators explicitly.
Note that the $BC_N$ Inozemtsev model is a universal completely integrable model of quantum mechanics with $B_N$ symmetry, which follows from the classification due to Ochiai, Oshima and Sekiguchi \cite{OOS}.
For the case $N=1$, finding eigenstates of the Hamiltonian is equivalent to solving the Heun equation \cite{OOS,Tak1}. In this sense, the $BC_N$ Inozemtsev model is a generalization of the Heun equation.

In this report, we are going to investigate eigenvalues and eigenstates of the $BC_N$ Inozemtsev model.

A possible approach is to use the quasi-exact solvability.
If the coupling constants $l$, $l_0$, $l_1$, $l_2$, $l_3$ satisfy some equation, the Hamiltonian $H$ (see (\ref{InoHam})) and the commuting operators of conserved quantities preserve some finite dimensional space of doubly periodic functions \cite{FGGRZ2,Takq}.
On the finite dimensional space, eigenvalues are calculated by solving the characteristic equation, and eigenfunctions are obtained by solving linear equations.
In this sense, a part of eigenvalues and eigenfunctions is obtained exactly and this is the reason to use the phrase ``quasi-exact solvability''.

Another approach is to use a method of perturbation.
By the trigonometric limit $p =\exp(\pi \sqrt{-1} \tau) \rightarrow 0$, the Hamiltonian $H$ of the $BC_N$ elliptic Inozemtsev model tends to the Hamiltonian of the $BC_N$ Calogero-Moser-Sutherland model $H_T$ (see (\ref{trigH})), and eigenvalues and eigenstates of the $BC_N$ Calogero-Moser-Sutherland model are known (see Proposition \ref{proptrig}). 

Based on the eigenstates for the case $p=0$, we can obtain eigenvalues and eigenstates of the $BC_N$ elliptic Inozemtsev model $(p\neq 0)$ as formal power series in $p$. This procedure is sometimes called the algorithm of perturbation.
% (see section \ref{sect:formalpert}).
Generally speaking, convergence of the formal power series obtained by perturbation is not guaranteed a priori, but for the case of the $BC_N$ elliptic Inozemtsev model, the convergence radius of the formal power series in $p$ is shown to be non-zero (see Corollary \ref{cor:conv}), and it is seen that this perturbation is holomorphic.
% perturbation and furthermore the holomorphic family of type (A). For definitions see the Kato's book \cite{Kat}.
As a result, real-holomorphy of the eigenvalues in $ p$ and the completeness of the eigenfunctions are shown.

In section \ref{sect:trig}, some results on the $BC_N$ trigonometric Calogero-Moser-Sutherland model are reviewed. In section \ref{sect:pert}, we present some propositions about the perturbation of the $BC_N$ elliptic Inozemtsev model from the trigonometric model. In section \ref{sect:comm}, we give some comments for future problems. 

%\section{$BC_N$ trogonometric Calogero-Moser-Sutherland model}
\section{Trigonometric limit} \label{sect:trig} $ $

In this section, we will consider the trigonometric limit $(\tau \rightarrow \sqrt{-1} \infty)$. and review some results on the trigonometric model. Assume $l, l_0, l_1\geq 0$ and set $p=\exp (\pi \sqrt{-1} \tau )$. Then $p\rightarrow 0$ as $\tau \rightarrow \sqrt{-1} \infty$.

If $p \rightarrow 0$, then 
$H \rightarrow H_{T} +C_T$,
where
\begin{eqnarray}
& H_T = -\sum_{j=1}^N\frac{\partial ^2}{\partial x_j^2} +\sum_{1\leq j<k\leq N} \left( \frac{2\pi ^2 l(l+1)}{\sin ^2\pi (x_j-x_k)} +\frac{2\pi ^2l(l+1)}{\sin ^2 \pi (x_j +x_k)} \right) & \label{trigH} \\
& \; \; \; \; \; + \sum_{j=1}^N \left( \frac{\pi^2 l_0(l_0+1)}{\sin ^2 \pi x_j} + \frac{\pi^2 l_1(l_1+1)}{\cos ^2 \pi x_j} \right) , & \nonumber
\end{eqnarray}
and $C_T= -\frac{N(N-1) \pi ^2}{3}l(l+1)-\frac{N\pi ^2}{3}\sum_{i=0}^3 l_i (l_i +1)$. The operator $H_T$ is nothing but the Hamiltonian of the $BC_N$ trigonometric Calogero-Moser-Sutherland model.

Now we solve the spectral problem for $H_T$ by using hypergeometric functions.
Set 
\begin{eqnarray}
&  \Phi_T(x)= \prod_{j=1}^N (\sin \pi x_j)^{l_0+1} (\cos \pi x_j)^{l_1+1} & \\
& \; \; \; \; \; \; \; \prod_{1\leq j_1<j_2\leq N} \left( \sin \pi (x_{j_1}-x_{j_2}) \sin \pi (x_{j_1 } +x_{j_2}) \right) ^{l+1}. & \nonumber 
\end{eqnarray}
Let  $D=\{ (x_1, \dots , x_N) \in \Rea ^N | 0 < x_1 \leq \dots \leq x_N<1 \}$. A function $f(x_1, \dots , x_N)$ is $W(B_N)$-invariant iff $f(x_1, \dots ,x_i, \dots ,x_j, \dots  x_N)=f(x_1, \dots ,x_j, \dots ,x_i, \dots  x_N) $ for all $i<j$ and $f(x_1, \dots ,x_i, \dots  x_N)=f(x_1, \dots , -x_i, \dots  x_N)$ for all $i$.
The Hilbert space ${\bf H}$ is defined by
\begin{equation}
{\bf H} = \left\{ f\!  : \Rea ^N \rightarrow \Cplx \mbox{ measurable} \left|
\begin{array}{l}
\int_D |f(x)| ^2 dx<+\infty, \\
\frac{f(x)}{\Phi_T(x)} \mbox{ is } W(B_N) \mbox{-invariant a.e. }x, \\
\frac{f(x+n)}{\Phi_T(x+n)} = \frac{f(x)}{\Phi_T(x)}\; ,\forall n\in \Zint^N  \mbox{ a.e. }x
\end{array}
\right. \right\} .
\label{Hilb1}
\end{equation}
\begin{prop} \label{proptrig}
Let ${\mathcal M}_N=\{ (\lambda _1, \dots ,\lambda _N ) \in \Zint^N | \lambda _1\geq \dots \geq \lambda _N \geq 0 \}$. There exists a complete orthonormal system $\{v_{\lambda } \}_{\lambda =(\lambda_1 ,\dots ,\lambda _N) \in {\mathcal M}_N}$ on the Hilbert space ${\bf H}$ such that 
$$
H_T v_{\lambda } = E_{\lambda } v_{\lambda },
$$
where $E_{\lambda } =\pi^2 \sum_{j=1}^N \left( 2\lambda _j +l_0+l_1+2+2(l+1)(N-j) \right) ^2$.
\end{prop}
Note that the function $v_{\lambda }$ is expressed as the product of the groundstate $ \Phi_T(x)$ and the $BC_N$ Jacobi polynomial $\psi _{\lambda }(z_1, \dots , z_N )$ $(z_j=\exp(2\pi \sqrt{-1} x_j ), \; j=1,\dots ,N )$. Essential selfadjointness of the operator $H_T$ on the Hilbert space ${\bf H}$ is obtained by applying Proposition \ref{proptrig}.

\section{Perturbation from the trigonometric model} \label{sect:pert}

We apply a method of perturbation and have an algorithm for obtaining eigenvalues and eigenfunctions as formal power series of $p$. The proofs of propositions in this report are obtained by imitating the proofs of the corresponding propositions in \cite{KT,Tak2}

Set $p=\exp(\pi \sqrt{-1} \tau) $. For the Hamiltonian of the $BC_N$ Inozemtsev model, we adopt a notation $H(p)$ instead of $H$. The operator $H(p)$ admits the following expansion:
\begin{equation}
H(p)(=H)=H_T +C_T+\sum_{k=1}^{\infty} V_k(x) p^k,
\label{Hamilp0}
\end{equation}
where $H_T$ is the Hamiltonian of trigonometric model defined in (\ref{trigH}),  $V_k(x)$ is some function in variables $x_1, \dots ,x_N$, and $C_T$ is a constant defined in section \ref{sect:trig}.

Based on the eigenvalues $E_{\lambda }$ $(\lambda \in {\mathcal M}_N )$ and the eigenfunctions $v_{\lambda }$ of the operator $H_T$, we determine eigenvalues $E_{\lambda }(p) = E_{\lambda }+C_T+\sum_{k=1}^{\infty} E_{\lambda }^{\{k\}}p^k$ and normalized eigenfunctions $v_{\lambda }(p)= v_{\lambda }+ \sum_{k=1}^{\infty} \sum_{\mu \in {\mathcal M}_N} c_{\lambda ,\mu }^{\{ k \} }v_{\mu }p^k$ of the operator $H(p)$
as formal power series in $p$.
In other words, we will find $E_{\lambda }(p) $ and $v_{\lambda }(p)$ that satisfy equations
\begin{eqnarray}
& H(p)v_{\lambda }(p)=(H_T +C_T +\sum_{k=1}^{\infty} V_k(x) p^k )v_{\lambda }(p) = E_{\lambda }(p)v_{\lambda }(p) ,\label{Hpertexp} &\\
& \langle v_{\lambda }(p) ,  v_{\lambda }(p) \rangle =1 , & \nonumber
\end{eqnarray}
as formal power series of $p$.

First we calculate coefficients $\sum_{\mu  \in {\mathcal M}_N} d_{\lambda ,\mu }^{\{k\}}v_{\mu }= V_k(x) v_{\lambda }$ $(k\in \Zint_{>0}, \; \lambda \in {\mathcal M}_N)$. 
%It is shown that for each $\lambda \in {\mathcal M}_N$ and $k\in \Zint_{>0}$, $d_{\lambda ,\mu }^{\{k\}}\neq 0$ for finitely many $\mu \in {\mathcal M}_N$.
Next we compute $E_{\lambda }^{\{k\}}$ and $ c_{\lambda ,\mu }^{\{k\}}$ for $k \geq 1$ and $\lambda , \mu \in {\mathcal M}_N$.  By comparing coefficients of $v_{\mu }p^k$ in (\ref{Hpertexp}), we obtain recursive relations for $E_{\lambda }^{\{k\}}$ and $c_{\lambda ,\mu }^{\{k\}}$. 

Now we present results which are obtained by applying the Kato-Rellich theory. We use definitions written in Kato's book \cite{Kat} freely.
It is shown that the operator $H(p)$ $(-1<p<1)$ is essentially selfadjoint on the Hilbert space ${\bf H}$, because the operator $\sum_{k=1}^{\infty} V_k(x) p^k$ is bounded.
Let $\tilde{H}(p)$ $(-1<p<1)$ be the unique extension of $H(p)$ to the selfadjoint operator.

\begin{prop} \label{prop:holA}
The operators $\tilde{H}(p)$ form a holomorphic family of type (A) for $-1<p<1$.
\end{prop}
%Let $\sigma (\tilde{H}(p))$ be the spectrum of the operator $\tilde{H}(p)$. Then the following proposition is proved:
\begin{prop} \label{prop:discr}
The spectrum $\sigma (\tilde{H}(p))$ contains only point spectra and it is discrete. The multiplicity of each eigenvalue is finite.
\end{prop}

Combining Theorem 3.9 in \cite[VII-\S 3.5]{Kat}, propositions in this report and the selfadjointness of $\tilde{H}(p)$, the following theorem is proved:
\begin{thm} \label{mainthmKato}
All eigenvalues of $\tilde{H}(p)$ $(-1<p<1)$ can be represented as $E_{\lambda } (p)$ $(\lambda \in {\mathcal M}_N)$, which is real-holomorphic in $p \in (-1,1)$ and $E_{\lambda } (0) =E_{\lambda } +C_T$.
% coincides with the eigenvalue of the trigonometric model which was obtained in section \ref{sect:trig} explicitly.
The eigenfunction $v_{\lambda } (p)$ of the eigenvalue $E_{\lambda } (p)$ is holomorphic in $p\in (-1,1)$ as an element in $L^2$-space, and the eigenvectors $v_{\lambda } (p)$ $(\lambda \in {\mathcal M}_N)$ form a complete orthonormal family on ${\bf H}$.
\end{thm}

As an application of the theorem, the convergence of formal power series of eigenvalues in the variable $p$ obtained by the algorithm of perturbation is shown.
\begin{cor} \label{cor:conv}
Let $E_{\lambda }(p)$ $({\lambda }\in {\mathcal M}_N)$ (resp. $v_{\lambda }(p)$) be the formal eigenvalue (resp. eigenfunction) of the Hamiltonian $H(p)$ defined by (\ref{Hpertexp}).
If $|p|$ is sufficiently small then the power series $E_{\lambda }(p)$ converges and as an element in $L^2$ space the power series $v_{\lambda }(p)$ converges.
\end{cor}

\section{Comments} \label{sect:comm}

%In \cite{KT}, the method of perturbation for the elliptic Calogero-Moser-Sutherland models from the trigonometric models is introduced.
In this report, holomorphy of perturbation for the Hamiltonian of the $BC_N$ Inozemtsev model from the trigonometric one is established.
Relationship between the perturbation and the complete integrability should be clarified. More precisely, holomorphy of perturbation for commuting operators of conserved quantities should be shown, although it is not succeed as of this writing.
For the elliptic Calogero-Moser-Sutherland model of type $A_N$, holomorphy of perturbation for commuting operators is shown in \cite{KT,Tak}.

In \cite{Tak1,Tak3}, Bethe Ansatz method for the $BC_1$ Inozemtzev model is proposed and some results on the finite gap integration for the $BC_1$ Inozemtzev model are established.
We hope that some progress on the Bethe Ansatz method or the finite gap integration for the $BC_N$ Inozemtzev model will be made.

%\section*{References}

\end{document}